\documentclass[aps,prc,reprint,superscriptaddress,longbibliography,nofootinbib,floatfix]{revtex4-2}

\usepackage{amsmath}
\usepackage{amssymb}
\usepackage{graphicx}
\usepackage{colortbl}
\usepackage{multirow}
\usepackage{xcolor}

\graphicspath{{./figures/}{./}}

\definecolor{rowgreen}{rgb}{0.85,1,0.85}
\definecolor{rowred}{rgb}{1,0.82,0.82}

\begin{document}

\title{Hint of universality in $\alpha$-cluster nuclei}
\author{Mario Gattobigio}
\affiliation{Institut de Physique de Nice (INPHYNI), CNRS UMR 7010, Universit\'e C\^ote d'Azur, Nice, France}

\date{\today}

\begin{abstract}
The empirical ground-state energies of the $\alpha$-cluster nuclei are placed on
the universal function that governs $N$-boson systems with a shallow state.  The
data fall on the finite-range-corrected universal curve, suggesting Efimov-type
universality.  No excited states, transition strengths, radii, or cluster
geometries enter the analysis; the nuclei are treated as $N$ structureless
bosons. A zero-shape $\alpha$-$\alpha$ Bargmann potential, describing the
$^{8}$Be resonance, provides the pairwise interaction with no shape freedom.
Simple three- and four-body forces, each a single Gaussian in the hyperradius,
are fitted to $^{12}$C and $^{16}$O; microscopic calculations for $^{20}$Ne and
$^{24}$Mg ($N=5,6$) follow as parameter-free predictions. A compact $0^+_2$
state near the Hoyle energy emerges in $^{12}$C without being targeted by the
fit.
\end{abstract}

\maketitle

\section{Introduction}
\label{sec:intro}

Light self-conjugate nuclei behave, to a remarkable degree, as if they were
built out of $\alpha$
particles~\cite{ikeda:1980_Prog.Theor.Phys.Supplement,freer:2018_Rev.Mod.Phys.,otsuka:2020_Rev.Mod.Phys.},
with the empirical organizing principle given by the Ikeda
diagram~\cite{ikeda:1968_Prog.Theor.Phys.Suppl.}.  Sixty years of
experiment~\cite{heydenburg:1956_Phys.Rev.,chien:1974_Phys.Rev.C,tilley:2004_Nucl.Phys.A}
and microscopic theory have filled this picture in considerable
detail~\cite{ali:1966_NuclearPhysics,afzal:1969_Rev.Mod.Phys.,buck:1977_NuclearPhysicsA,oertzen:2006_Eur.Phys.J.A,%
freer:2018_Rev.Mod.Phys.,nakamura:2025_Eur.Phys.J.A},
from antisymmetrized molecular dynamics and the $\alpha$-condensate wave
function~\cite{tohsaki:2001_Phys.Rev.Lett.,funaki:2010_J.Phys.GNucl.Part.Phys.}
to \emph{ab initio} lattice~\cite{epelbaum:2011_Phys.Rev.Lett.}, zero-range 
effective field theory (EFT)~\cite{higa:2008_Nucl.Phys.A,schmickler:2019_PhysicsLettersB}, and shell-model
calculations~\cite{otsuka:2022_NatCommun} that resolve the internal geometry of
individual states.

This paper asks a deliberately narrower question. From the whole
$\alpha$-cluster spectrum we consider only the
ground-state energy measured from the $N\alpha$ decay threshold, which is the value
displayed by the Ikeda diagram itself. We discard everything else: no excited
states, no transition strengths, no radii, no geometrical arrangement of the
clusters. The nuclei are treated as $N$ structureless bosons, which is
manifestly not what they are. The question is whether that one number, across
the empirical series of the $\alpha$ clusters, follows a universal law.

The universality explored here is that of
Efimov~\cite{efimov:1970_Phys.Lett.B,efimov:1973_Nucl.Phys.A}, which has been extensively
reviewed in Refs.~\cite{braaten:2006_PhysicsReports,naidon:2017_Rep.Prog.Phys.}.
In this work, we use the linear representation of the finite-range-corrected
universal Efimov law for $N$-boson
systems with a shallow state~\cite{gattobigio:2014_Phys.Rev.A,kievsky:2014_Phys.Rev.A,%
kievsky:2015_Phys.Rev.A}.
We use the values extracted from the Ikeda thresholds, normalized by the 
energy of the $^8$Be resonance, to evaluate the universal function, 
showing that the points fall cleanly on the universal curve.

The rest of the paper investigates which interaction reproduces that line.
Section~\ref{sec:universalfunction} establishes the linearity and shows that no
member of a one-parameter family of two-body potentials can account for it: the
$\alpha$-cluster ground states are not a two-body problem.
Section~\ref{sec:twobody} fixes the pairwise input with as little freedom as
possible, by requiring the $\alpha$-$\alpha$ interaction to be a zero-shape
$S$-matrix~\cite{bargmann:1949_Rev.Mod.Phys.,recchia:2022_Phys.Rev.A} anchored
to the $^{8}$Be resonance, so that whatever is missing at $N>2$ cannot have
been hidden in a shape parameter. Section~\ref{sec:manybody} adds simple
three- and four-body forces, fits their two strengths
to $^{12}$C and $^{16}$O, and performs calculations up to $N=6$, leaving
$^{20}$Ne and $^{24}$Mg as parameter-free predictions.

One result was not sought. Nothing in this construction asks for a $3\alpha$
excited state. Nevertheless, once the
three-body force is switched on, $^{12}$C acquires a $0^+_2$ state just above the
$3\alpha$ threshold, close to where the Hoyle
state~\cite{hoyle:1954_Astrophys.J.Suppl.Ser.,freer:2014_Prog.Part.Nucl.Phys.}
is observed. We report it in Sec.~\ref{sec:hoyle} for what it is -- a byproduct
of a model designed for another purpose -- and not as a
full description of that state. Its width, its monopole strength, and the geometry of
  its wave function are outside the scope of this work.
\section{Universality of the Ikeda thresholds}
\label{sec:universalfunction}

We use the 
energies of the $\alpha$-cluster nuclei given in the Ikeda
diagram~\cite{ikeda:1968_Prog.Theor.Phys.Suppl.,nakamura:2025_Eur.Phys.J.A}
with respect to their $N\alpha$-decay thresholds. The quantity that
organizes them is the ground-state energy measured from that threshold
\begin{equation}
  E_N = N\times B(^{4}\mathrm{He}) - B_N\,,
  \label{eq:deltab}
\end{equation}
where $B_N$ is the binding energy of the nucleus made of $N$ $\alpha$-particles;
in this convention $E_N<0$ for a state bound below the threshold.
These numbers are listed in Table~\ref{tab:ikeda}.

The two-body member of the family is special: $^{8}$Be is not bound at all, but
a resonance at $E_2=+91.84$~keV above the $2\alpha$
threshold~\cite{benn:1967_Nucl.Phys.A,tilley:2004_Nucl.Phys.A}. It is
nevertheless the two-body scale of the problem, and everything below is
measured against its magnitude. Using that magnitude is legitimate because
what the zero-shape universal law asks for is a two-body $S$-matrix pole anomalously
close to threshold on the scale set by the range of the force -- what can be
called a shallow state, either bound or virtual.

For $N$ identical bosons near the unitary limit, the Efimov
spectrum~\cite{efimov:1970_Phys.Lett.B,efimov:1973_Nucl.Phys.A} generalizes to
the $N$-body law~\cite{gattobigio:2014_Phys.Rev.A}
\begin{equation}
  \frac{|E_N|}{|E_2|}=\tan^2\xi\,,\qquad
  \kappa_N a_B+\Gamma_N=\frac{e^{-\Delta(\xi)/2s_0}}{\cos\xi}\equiv y(\xi)\,,
  \label{eq:ynxi}
\end{equation}
where $\kappa_N=\sqrt{m|E_N^*|}/\hbar$, $E_N^*$ is the $N$-body energy at the unitary point,
and the two-body binding length $a_B$ is defined by
$E_2=\hbar^2/(m a_B^2)$. As we are doing calculations for $\alpha$ clusters, we use 
$\hbar^2/m = 10.446465\;\text{MeV fm}^2$ for the mass. The angle $\xi$ carries all the $N$ dependence of the
energies, and $\Delta(\xi)$ is the Efimov universal function with
$s_0=1.00624$~\cite{gattobigio:2019_Few-BodySyst.,gattobigio:2014_Phys.Rev.A}.
The two numbers $\kappa_N$ and $\Gamma_N$ are the
only $N$-dependent input, and for a fixed two-body potential both are found to
be linear in $N$~\cite{kievsky:2014_Phys.Rev.A,recchia:2022_Few-BodySyst},
\begin{equation}
  \kappa_N = A_1+A_2N\,,\qquad \Gamma_N = C_1+C_2N\,,
  \label{eq:kappagamma}
\end{equation}
which makes $y$ itself an approximately linear function of $N$.
The linearity of $\kappa_N$ is not an artifact of the finite-range potentials
used in Refs.~\cite{kievsky:2014_Phys.Rev.A,recchia:2022_Few-BodySyst}: the same
behavior had already been predicted for the $N$-boson ground state in the
zero-range limit. From the log-normal statistics of the two-particle correlator
at unitarity, Ref.~\cite{nicholson:2012_Phys.Rev.Lett.} obtains
$|E^*_N|=\tfrac{1}{8}N(N-2)\,|E^*_4|$ for even $N$, while the leading-order
EFT calculations of Ref.~\cite{bazak:2016_Phys.Rev.A},
converged to zero range, give $|E^*_N|/|E^*_3|\simeq(N-2)^2$. In both cases
$|E^*_N|\propto N^2$ up to an offset, that is, a $\kappa_N$ linear in $N$.

\begin{figure}[t]
  \centering
  \includegraphics[width=0.9\linewidth]{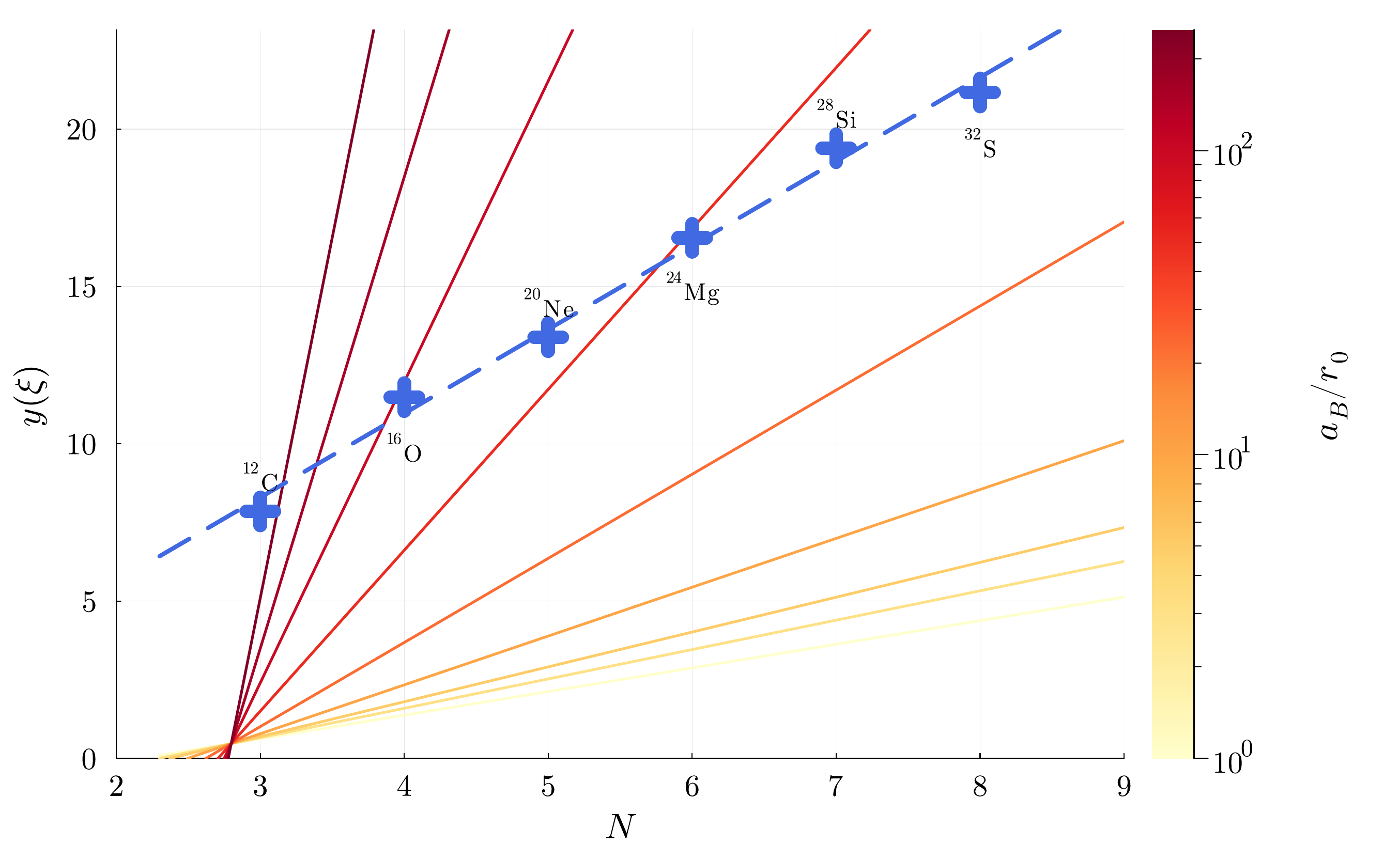}
  \caption{The universal function $y$ against the number of clusters $N$.
  Crosses: the $\alpha$-cluster thresholds of Table~\ref{tab:ikeda}, with the
  straight line of Eq.~\eqref{eq:ylinear} through them. Solid lines: $N$-boson
  calculations with a two-body Gaussian
  potential $V_0\,\exp{(-r^2/r_0^2)}$~\cite{kievsky:2014_Phys.Rev.A}, colored by the ratio $a_B/r_0$
  of the two-body scale to the potential range. Each family member is linear
  in $N$.}
  \label{fig:ynxi-real}
\end{figure}

In Fig.~\ref{fig:ynxi-real} we report the Ikeda data in blue, and 
we observe a linear behavior with the fit given by Eq.~(\ref{eq:ylinear})
\begin{equation}
  y(N)\simeq 0.29 + 2.67\,N\,,
  \label{eq:ylinear}
\end{equation}
as already predicted for bosons in Ref.~\cite{gattobigio:2014_Phys.Rev.A,kievsky:2014_Phys.Rev.A,recchia:2022_Few-BodySyst}. 
Equation~\eqref{eq:ylinear} suggests that the Ikeda thresholds are universal in
the Efimov sense.

\begin{table}[t]
  \caption{The $\alpha$-cluster thresholds and their position on the
  universal function. $E_N$ is the ground-state energy relative to the
  $N\alpha$ threshold, that is, the number in the Ikeda
  diagram~\cite{ikeda:1968_Prog.Theor.Phys.Suppl.,nakamura:2025_Eur.Phys.J.A}.
  The angle is
  $\xi_N=-\arctan\sqrt{|E_N/E_2|}$ with $E_2=91.84$~keV, the $^{8}$Be
  resonance energy; $y(\xi_N)$ follows from Eq.~\eqref{eq:ynxi}
  using the parametrization given in Ref.~\cite{gattobigio:2019_Few-BodySyst.}.}
  \label{tab:ikeda}
  \begin{ruledtabular}
  \begin{tabular}{clcccc}
    $N$ & nucleus & $E_N$ (MeV) & $|E_N/E_2|$ & $\xi_N/\pi$ & $y(\xi_N)$ \\
    \hline
    2 & $^8$Be    & $+0.092$  & -     &   -      &   -   \\
    3 & $^{12}$C  & $-7.27$  &  79.0 & $-0.4643$ &  7.85 \\
    4 & $^{16}$O  & $-14.44$ & 157.0 & $-0.4747$ & 11.48 \\
    5 & $^{20}$Ne & $-19.17$ & 208.4 & $-0.4780$ & 13.39 \\
    6 & $^{24}$Mg & $-28.48$ & 309.6 & $-0.4819$ & 16.54 \\
    7 & $^{28}$Si & $-38.46$ & 418.0 & $-0.4844$ & 19.39 \\
    8 & $^{32}$S  & $-45.41$ & 493.6 & $-0.4857$ & 21.16 \\
  \end{tabular}
  \end{ruledtabular}
\end{table}

The question is then whether a two-body force can produce that line;
Fig.~\ref{fig:ynxi-real} answers it. Plotted against the Ikeda data is a family
of $N$-boson calculations with a two-body Gaussian
potential, $V_0
\exp(-r^2/r_0^2)$~\cite{kievsky:2014_Phys.Rev.A,recchia:2022_Few-BodySyst}, where 
each member is labeled by the ratio $a_B/r_0$ of the two-body scale to the
potential range, using a heat map ranging from yellow to dark red. Every member is
itself linear in $N$, so the functional form is right; but it is impossible to
reproduce the Ikeda data with any single member of this family of two-body
potentials.
This is the sense in which the $\alpha$-cluster spectrum is
not a two-body problem, and it is a statement about the family as a whole,
not about a particular fit.

Since a straight line is fixed by two of its points, the natural remedy is to
let the data fix it: we use $N=3$ and $N=4$ to calibrate genuine three- and
four-body forces, and then see whether $N=5$ and $N=6$ follow. That is the
strategy of Sec.~\ref{sec:manybody}. First, however, the pairwise input has
to be fixed in a way that leaves as little freedom as possible, so that
whatever the many-body force has to supply is not simply making up for an
arbitrary choice of the two-body potential $V_2$;
this is achieved by forcing the two-body short-range potential to be a member
of the zero-shape universal family~\cite{recchia:2022_Phys.Rev.A,recchia:2022_Few-BodySyst}.
\section{The two-body interaction}
\label{sec:twobody}

\subsection{The zero-shape universality}
\label{sec:smatrix}

The two-body Schr\"odinger problem is equivalent to its $S$-matrix: specifying
the $S$-matrix is the same as specifying a two-body potential. For the moment we concentrate
only on the short-range potential. The 
simplest $S$-matrix describing a single bound state is
\begin{equation}
  S(k) = \frac{k+i/a_B}{k-i/a_B}\,\frac{k+i/r_B}{k-i/r_B}\,,
  \label{eq:Smatrix}
\end{equation}
with a physical pole at $k=i/a_B$, which gives the two-body energy, and a
second pole at $k=i/r_B$. The latter carries no state; $r_B$ instead sets the
scale of the underlying theory. Writing $S(k)=1-i(km/2\pi)T(k)$ and matching
to the effective-range expansion
\begin{equation}
  T(k) = -\frac{4\pi}{m}
    \left(-\frac1a+\frac12 r_e k^2 + v_2 k^4+\cdots-ik\right)^{-1}
  \label{eq:ere}
\end{equation}
one identifies Eq.~\eqref{eq:Smatrix} as precisely the point at which all shape
parameters vanish, $v_n=0$, and fixes the remaining low-energy constants
through
\begin{equation}
  r_B = a - a_B\,, \qquad r_e\,a = 2\,r_B\,a_B\,.
  \label{eq:zeroshape-relations}
\end{equation}
Equation~\eqref{eq:Smatrix} gives the maximally constrained pairwise
S-matrix consistent with a single bound state: beyond the two scales $a_B$
and $r_B$ there is nothing left to choose on shell. Equation~(\ref{eq:zeroshape-relations}) 
defines a curve in the $(r_e/a,\,r_e/a_B)$ plane on which any zero-shape
system must lie; this is the blue curve in Fig.~\ref{fig:zeroshape}, where we have also
placed real two-body systems. We use experimental data where possible, and results
obtained using ab initio potentials in other
cases~\cite{kievsky:2015_Phys.Rev.A,alvarez-rodriguez:2016_Phys.Rev.A,kievsky:2021_Annu.Rev.Nucl.Part.Sci.}.
Moreover, we add a recent experimental measurement in the $pp$ system without
Coulomb, where the zero-shape hypothesis has been
tested~\cite{tumino:2023_Commun.Phys.}.

\subsection{The Bargmann family and its off-shell parameter}
\label{sec:bargmann}

A one-parameter family of local potentials -- the Bargmann
potentials~\cite{bargmann:1949_Rev.Mod.Phys.} -- realizes Eq.~\eqref{eq:Smatrix} exactly.
Writing
\begin{equation}
  V_c(r) = -\frac{\hbar^2}{m}\,4\kappa\,\frac{d}{dr}\!\left[
    \frac{\sinh(\lambda r/2)\,g_c(\kappa,r)}{g_c(\mu_+,r)-g_c(-\mu_-,r)}\right],
  \label{eq:bargmann}
\end{equation}
with
\begin{equation}
  g_c(k,r)=\frac{e^{-kr}+c\sinh(kr)}{k}\,,
  \label{eq:gc}
\end{equation}
and
\begin{equation}
  \kappa=\frac{\lambda(\beta-1)}{2(\beta+1)}\,,\qquad
  \mu_+=\frac{\lambda\beta}{\beta+1}\,,\qquad
  \mu_-=\frac{\lambda}{\beta+1}\,,
  \label{eq:bargmann-scales}
\end{equation}
the on-shell scales are fixed by $\beta$ and $\lambda$, while $c$ is a
pure off-shell parameter: every member of the family has the same $S$-matrix
given in Eq.~(\ref{eq:Smatrix}).
A special case is $c=2$, and it is known as the Eckart potential
\begin{equation}
  V_{\rm Eckart}(r) = -\frac{\hbar^2}{m}\,
    \frac{2\beta\lambda^2 e^{-\lambda r}}{\left(1+\beta e^{-\lambda r}\right)^2}\,.
  \label{eq:eckart}
\end{equation}

The two parameters $\beta$ and $\lambda$ carry all the on-shell
information, and the low-energy constants follow from them in closed form.
The physical pole of Eq.~\eqref{eq:Smatrix} is the $\kappa$ of
Eq.~\eqref{eq:bargmann-scales}, so that $a_B=1/\kappa$, and
\begin{equation}
  a_B = \frac{2(\beta+1)}{\lambda(\beta-1)}\,,\qquad
  r_B = \frac{2}{\lambda}\,,
  \label{eq:bargmann-aB-rB}
\end{equation}
\begin{equation}
  a = a_B + r_B = \frac{4\beta}{\lambda(\beta-1)}\,,\qquad
  r_e = \frac{2(\beta+1)}{\lambda\beta}\,,
  \label{eq:bargmann-a-re}
\end{equation}
with $B_2=\hbar^2\kappa^2/m$. The spurious pole is thus set by $\lambda$
alone, $r_B=2/\lambda$: $\lambda$ fixes the scale of the underlying theory,
and $\beta$ is the one remaining dimensionless parameter. Note that
Eqs.~\eqref{eq:bargmann-aB-rB} and \eqref{eq:bargmann-a-re} satisfy
\begin{equation}
  r_e\,a = 2\,r_B\,a_B = \frac{8(\beta+1)}{\lambda^2(\beta-1)}
  \label{eq:bargmann-zeroshape}
\end{equation}
identically, for every $\beta$ and $\lambda$. This is
Eq.~\eqref{eq:zeroshape-relations} again, and it is the precise sense in
which the family sits on the zero-shape curve by construction.

The off-shell parameter $c$, besides being visible in many-body calculations, determines the
root-mean-square radius, $r_{\rm rms}$, which depends on the wave function. 
Table~\ref{tab:offshell} makes the off-shell character of $c$ explicit:
at fixed $\beta$ and $\lambda$ we report the two-body binding energy and
$r_{\rm rms}$. This freedom will matter once we add the Coulomb interaction
between $\alpha$ particles.

\begin{table}[t]
  \caption{The Bargmann family at fixed on-shell data ($\beta=2.4386$,
  $\lambda=2.2066$~fm$^{-1}$), varying only the off-shell parameter $c$.
  The two-body binding energy $B_2=\hbar^2\kappa^2/m$, here without the
  Coulomb interaction, is unchanged; the size of the bound state is not. The
  value used throughout this work is $c=0.535$.}
  \label{tab:offshell}
  \begin{ruledtabular}
  \begin{tabular}{lccccc}
    $c$ & $0.3$ & $0.535$ & $1.0$ & $2.0$ (Eckart) & $5.0$ \\
    \hline
    $B_2$ (MeV)        & $2.2257$ & $2.2257$ & $2.2257$ & $2.2257$ & $2.2257$ \\
    $r_{\rm rms}$ (fm) & $1.854$   & $1.612$   & $1.371$   & $1.131$   & $0.862$ \\
  \end{tabular}
  \end{ruledtabular}
\end{table}

\subsection{Anchoring to the $\alpha$-$\alpha$ $^8$Be resonance}
\label{sec:alphaalpha}

The $\alpha$-$\alpha$ interaction is taken as the short-range
Bargmann potential plus the point-like Coulomb $\alpha$-$\alpha$
potential
\begin{equation}
V_2(r) = V_c(r) + \frac{4e^2}{r} \,,
\label{eq:v2}
\end{equation}
with $e^2=1.43996$~MeV~fm.
The three Bargmann parameters are fixed by requiring that the $^{8}$Be resonance,
$E_r=91.84$~keV and
$\Gamma=5.57$~eV~\cite{benn:1967_Nucl.Phys.A,tilley:2004_Nucl.Phys.A}, be
reproduced.
This gives
\begin{equation}
  \beta = 2.43860\,,\quad \lambda = 2.2066~\mathrm{fm}^{-1}\,,\quad
  c = 0.53530\,,
  \label{eq:params}
\end{equation}
and shows the important role of the off-shell parameter $c$.

\begin{figure}[t]
  \centering
  \includegraphics[width=0.9\linewidth]{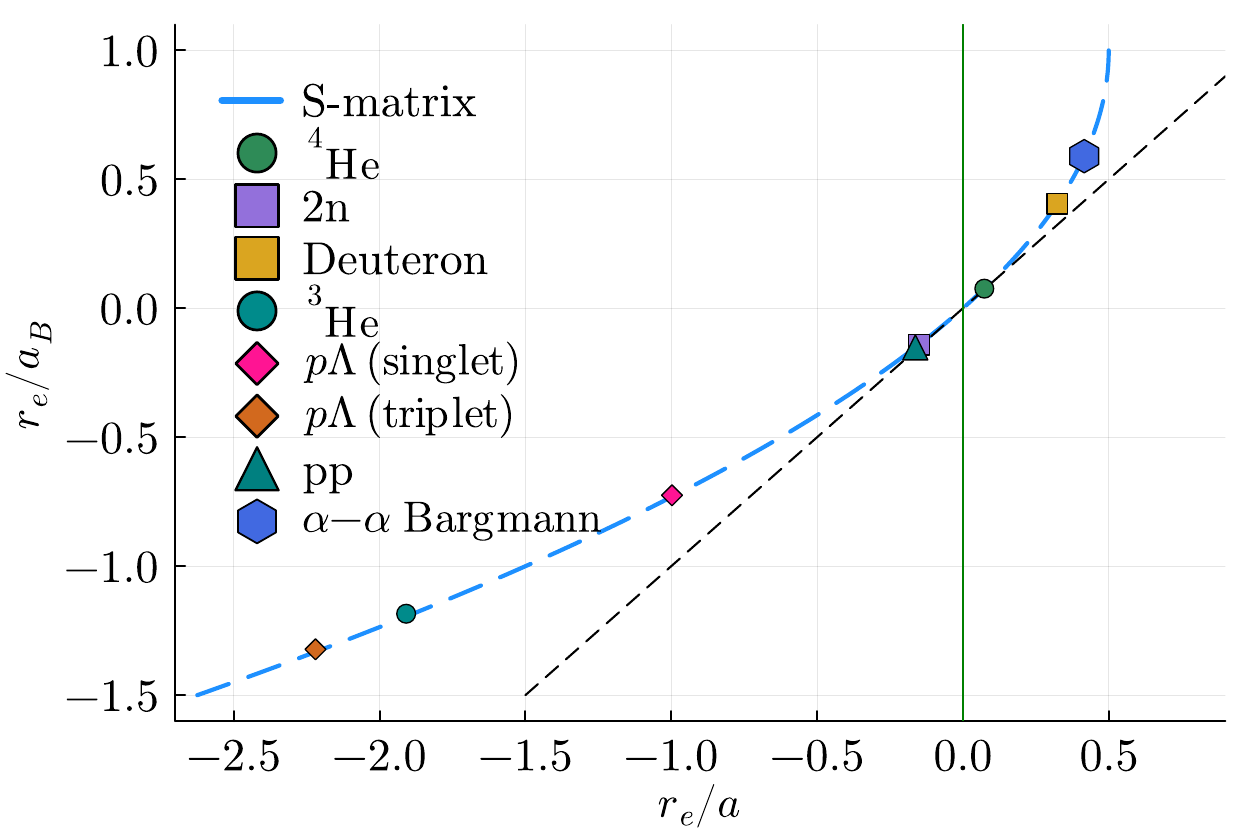}
  \caption{The zero-shape universality plane. The curve is the exact
  $S$-matrix relation $r_e a = 2 r_B a_B$, Eq.~\eqref{eq:zeroshape-relations};
  points are real two-body systems~\cite{kievsky:2015_Phys.Rev.A,alvarez-rodriguez:2016_Phys.Rev.A,tumino:2023_Commun.Phys.}.
  The $\alpha$-$\alpha$ ($^{8}$Be) point is the Bargmann potential of
  Eq.~\eqref{eq:params}; it lies on the curve by construction, the shape
  parameter vanishing identically for the whole Bargmann family. The
  straight dashed line is $r_e/a=r_e/a_B$.}
  \label{fig:zeroshape}
\end{figure}

Two remarks are in order. First, holding both $E_r$ and $\Gamma$ exactly is a
strong constraint, and it is known to sit in mild tension with the
$\alpha$-$\alpha$ $s$-wave scattering data: relaxing either one improves the
phase-shift $\chi^2$ appreciably, an effect already noted in the halo-EFT
treatment of Ref.~\cite{higa:2008_Nucl.Phys.A}. We keep the resonance exact, since it is
the two-body datum that the whole $N$-body construction is anchored to.
Second, and this is the point of choosing the Bargmann form, the resulting
potential lies \emph{on} the zero-shape curve by construction
(Fig.~\ref{fig:zeroshape}), at $r_e/a=0.4159$ and $r_e/a_B=0.5899$, which
satisfy Eq.~\eqref{eq:zeroshape-relations} identically. The
$\alpha$-$\alpha$ interaction is therefore described here with no shape
freedom at all: whatever the $N>2$ systems turn out to need, it cannot have
been absorbed into a shape parameter of the pairwise force.

\subsection{Results with pairwise force alone}
\label{sec:twobodyonly}

\begin{figure}[t]
  \centering
  \includegraphics[width=0.95\linewidth]{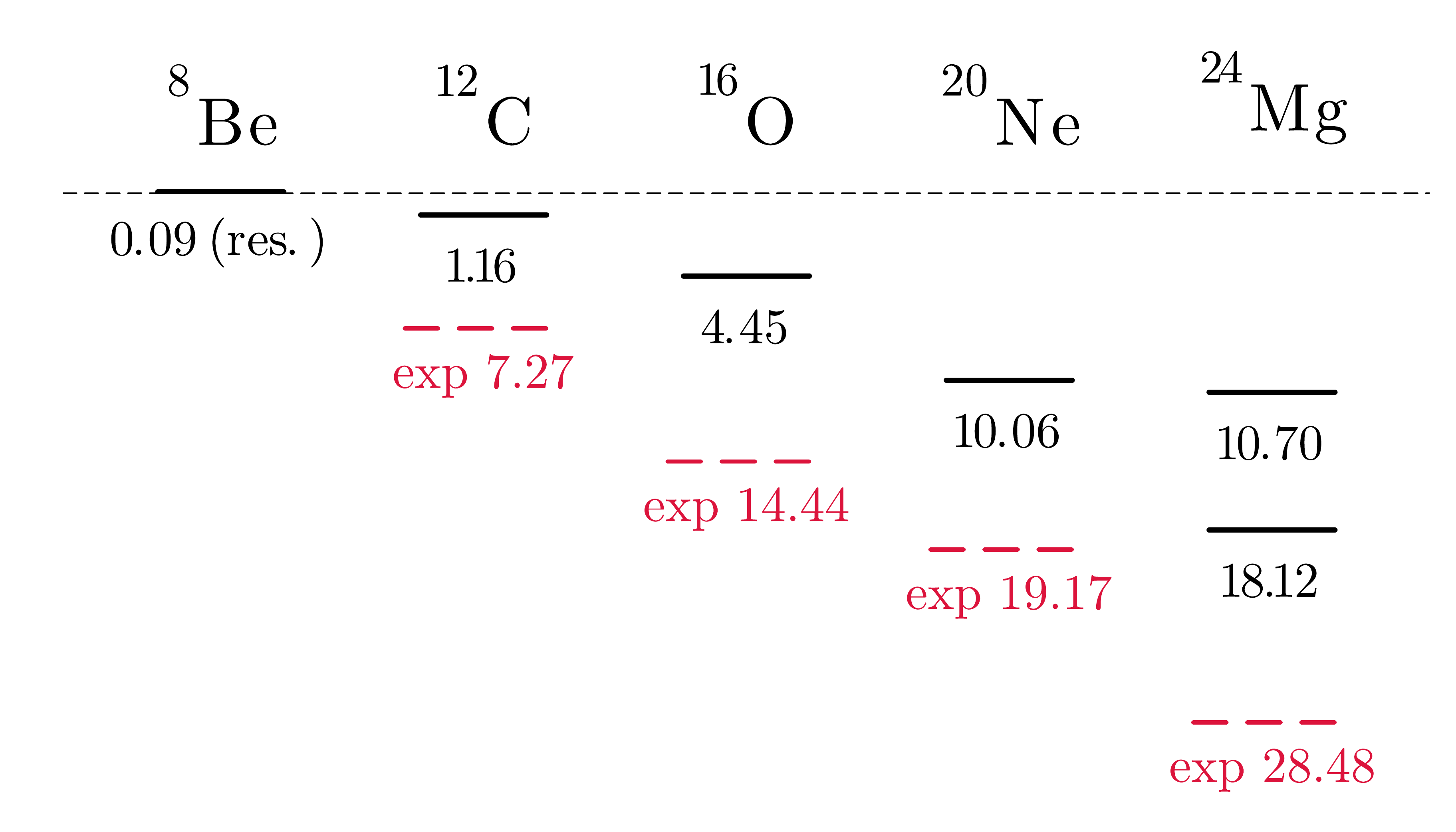}
  \caption{$N=2\ldots6$ spectrum, not to scale, from the Bargmann $\alpha$-$\alpha$ force
  alone, with Coulomb included; energies in MeV relative to the $N\alpha$
  threshold, experimental values dashed and in red. $^{8}$Be is reproduced by
  construction and is a resonance; every heavier system is underbound.}
  \label{fig:twobody-only-levels}
\end{figure}

With $V_2$ fixed, the $N$-body ground states were computed with the
stochastic variational method (SVM) on a correlated Gaussian
basis~\cite{varga:1995_Phys.Rev.C,suzuki:1998_}. Table~\ref{tab:twobodyonly} and
Fig.~\ref{fig:twobody-only-levels} give the results for $N=3\ldots6$, with and
without the Coulomb interaction.

\begin{table}[t]
  \caption{$N\alpha$ spectrum from the Bargmann two-body force alone
  (Eq.~\eqref{eq:params}), in MeV relative to the $N\alpha$ threshold, with and
  without the Coulomb interaction. $E_0$ is the ground state, $E_1$ the first
  excited state; the $N=3$ excited state is unbound once Coulomb is included.
  The last column is the experimental $E_N$ of Table~\ref{tab:ikeda}. Every
  charged system is underbound, and the deficit grows with $N$.}
  \label{tab:twobodyonly}
  \begin{ruledtabular}
  \begin{tabular}{clcccccc}
    & & \multicolumn{2}{c}{no Coulomb} & \multicolumn{2}{c}{with Coulomb} & \\
    $N$ & nucleus & $E_0$ & $E_1$ & $E_0$ & $E_1$ & exp.\ $E_N$ \\
    \hline
    3 & $^{12}$C  & $-9.67$  & $-3.54$  & $-1.16$  & ---      & $-7.27$  \\
    4 & $^{16}$O  & $-23.76$ & $-14.98$ & $-4.45$  & $-0.71$  & $-14.44$ \\
    5 & $^{20}$Ne & $-45.50$ & $-34.34$ & $-10.06$ & $-3.88$  & $-19.17$ \\
    6 & $^{24}$Mg & $-75.73$ & $-62.16$ & $-18.12$ & $-10.70$ & $-28.48$ \\
  \end{tabular}
  \end{ruledtabular}
\end{table}

The pattern is unambiguous: the two-body force underbinds the $\alpha$-cluster nuclei.
A pairwise force tuned exactly at
$N=2$ does not carry over to $N>2$. That is the gap the
next section is meant to fill.

\section{A minimal three- and four-body force}
\label{sec:manybody}

\subsection{The ansatz}
\label{sec:ansatz}

The nuclei are treated as systems of $N$ structureless bosonic
$\alpha$ particles, governed by the potential
\begin{equation}
  V =
     \sum_{i<j}V_2(r_{ij})
    + \sum_{i<j<k}V_3(\rho_3)
    + \sum_{i<j<k<l}V_4(\rho_4)\,.
  \label{eq:hamiltonian}
\end{equation}
The
three- and four-body terms are the most economical local forces one can
write: a single Gaussian in the hyperradius of the triplet and of the
quadruplet, $\rho_n^2=\sum_{\rm pairs}r_{ij}^2$,
\begin{equation}
  V_3 = W_3\,e^{-\rho_3^2/\rho_0^2}\,,\qquad
  V_4 = W_4\,e^{-\rho_4^2/\rho_0^2}\,.
  \label{eq:manybodyforce}
\end{equation}
A restrictive assumption is that the two forces 
share a \emph{single common range}
$\rho_0$. The construction is in the same spirit as the same-range many-body
forces used to study a four-body limit cycle at
unitarity~\cite{frederico:2023_Phys.Rev.A}, and as hyperradial energy-density-functional
constructions~\cite{kievsky:2021_Phys.Rev.A,orlandini:2025_J.Subat.Part.Cosmol.}; here the force 
describes real nuclear data, away from the strict unitary limit.

At each fixed $\rho_0$ the two strengths are determined by reproducing
$E(^{12}\mathrm{C})=-7.27$~MeV and  $E(^{16}\mathrm{O})=-14.44$~MeV.  Only these
two numbers enter. The $N=5$ and $N=6$ systems receive no further input, and
their energies are parameter-free predictions.

\subsection{The scan in $\rho_0$}
\label{sec:results}

\begin{figure}[t]
  \centering
  \includegraphics[width=0.95\linewidth]{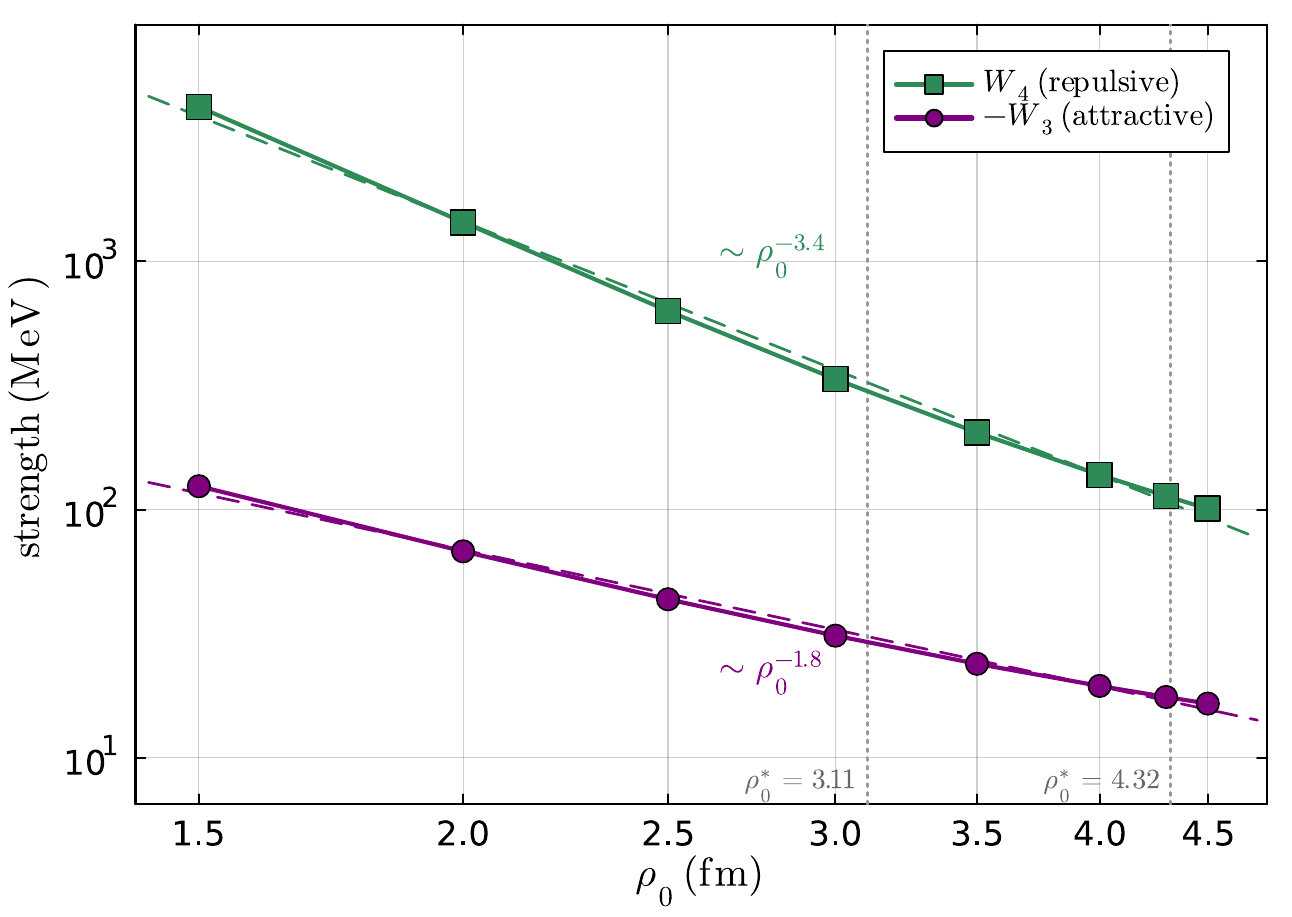}
  \caption{The three- and four-body strengths along the scan. At each
  $\rho_0$, $W_3$ and $W_4$ are refitted so that $^{12}$C and $^{16}$O stay
  at their experimental energies; plotted are $|W_3|$ and $W_4$, both
  positive, on logarithmic axes. $W_3$ is attractive and $W_4$ repulsive at
  every point, neither sign being imposed by the fit. Dashed lines are the
  power laws of Eq.~\eqref{eq:strengthscaling}. The dotted verticals mark
  the two ranges $\rho_0^\ast$ selected in Fig.~\ref{fig:scan} by $^{20}$Ne
  and by $^{24}$Mg.}
  \label{fig:strengths}
\end{figure}

In Fig.~\ref{fig:strengths} we show the strengths fitted at each point of
the scan in $\rho_0$ so as to reproduce $^{12}$C and $^{16}$O. Note first that $W_3$
comes out \emph{attractive} throughout and $W_4$ \emph{repulsive}. 

\begin{figure}[t]
  \centering
  \includegraphics[width=0.95\linewidth]{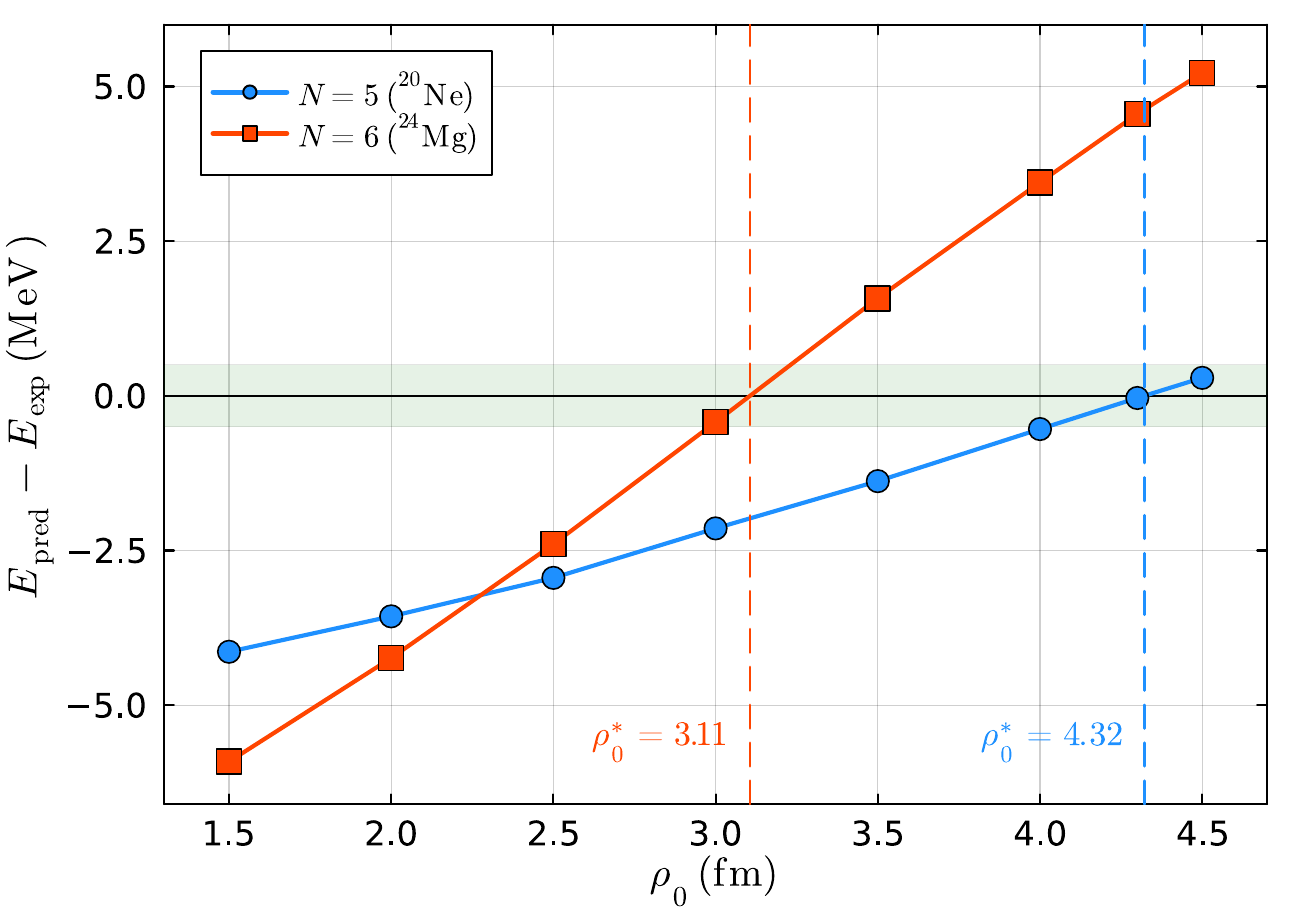}
  \caption{Deviation of the parameter-free $N=5$ and
  $N=6$ predictions from experiment as a function of the common range
  $\rho_0$ of the three- and four-body force.
  $N=3$ and $N=4$ are refitted at every point and would lie on the zero line
  by construction. Both curves are monotonic and both cross zero, but at
  different ranges: $\rho_0^\ast=4.32$~fm for $^{20}$Ne and
  $\rho_0^\ast=3.11$~fm for $^{24}$Mg. The shaded band marks $\pm0.5$~MeV.}
  \label{fig:scan}
\end{figure}

Both strengths then vary strongly across the scan, by a factor $7.5$ for
$W_3$ and a factor $42$ for $W_4$; a Gaussian with a smaller range has to
be deeper to do the same work. The variation is close to a power law in
the range,
\begin{equation}
  |W_3| \sim \rho_0^{-1.83}\,,\qquad W_4 \sim \rho_0^{-3.39}\,,
  \label{eq:strengthscaling}
\end{equation}
holding to within $6\%$ and $9\%$, respectively.
In Fig.~\ref{fig:scan} we report the deviation of the parameter-free
$N=5$ and $N=6$ predictions from the experimental values.

\begin{figure}[t]
  \centering
  \includegraphics[width=0.95\linewidth]{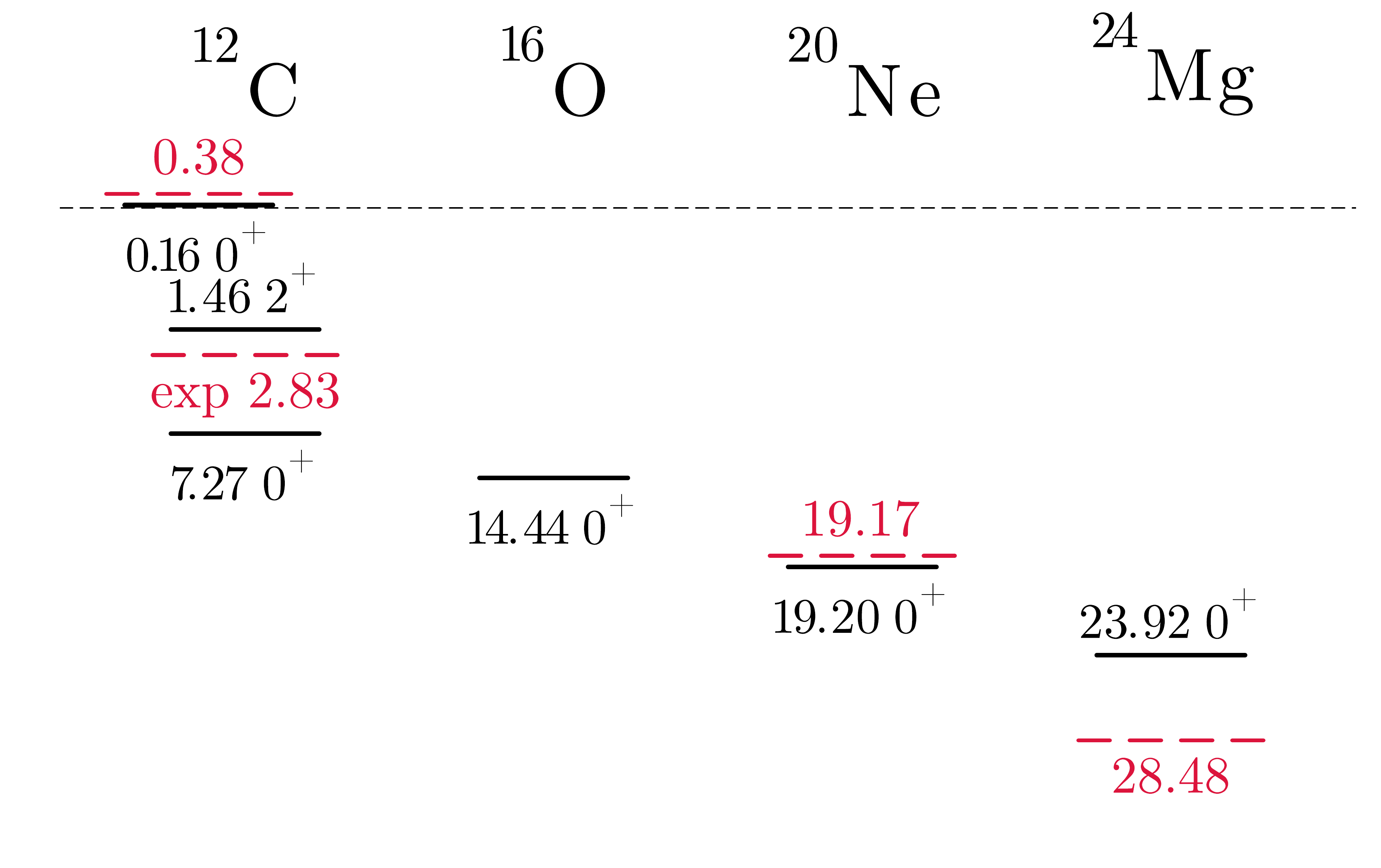}\\[2ex]
  \includegraphics[width=0.95\linewidth]{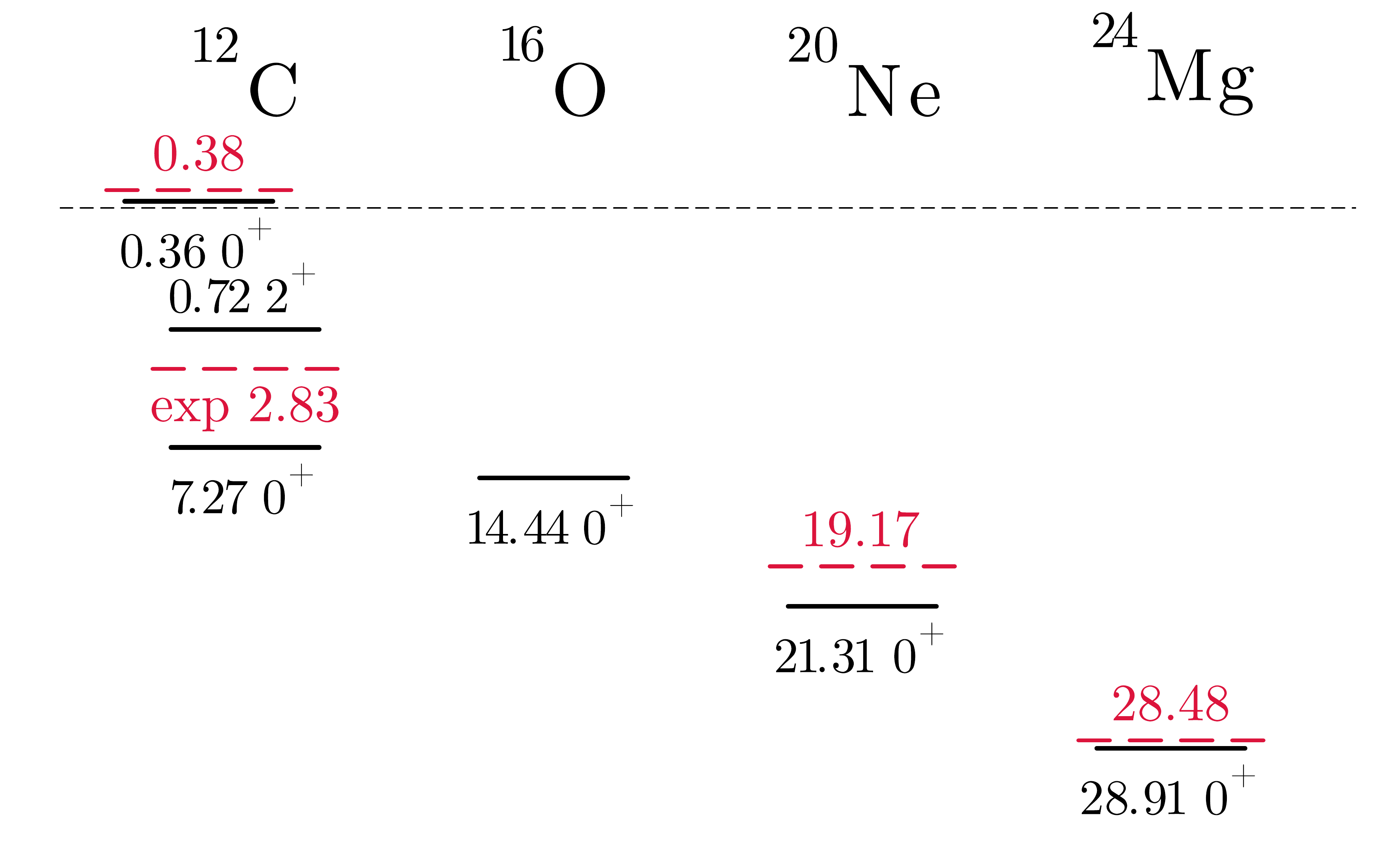}
  \caption{The two best points of the scan, as level schemes, not to scale; energies in MeV
  relative to the $N\alpha$ threshold, while the experimental data are dashed. Upper panel:
  $\rho_0=4.3$~fm, the range that reproduces $^{20}$Ne. Lower panel:
  $\rho_0=3.0$~fm, the range that reproduces $^{24}$Mg. In both, $^{12}$C and
  $^{16}$O are fitted and $^{20}$Ne and $^{24}$Mg are predicted; the $^{12}$C
  column is drawn schematically, its three states being far too close
  together to share the vertical scale of the other columns.}
  \label{fig:levels}
\end{figure}

Both prediction curves are monotonic in $\rho_0$ and both cross zero, which
is already a non-trivial statement: within this one-parameter family of
many-body forces there exists a range that reproduces $^{20}$Ne exactly, and
a range that reproduces $^{24}$Mg exactly. Linear interpolation between the
scan points locates the two crossings at
\begin{equation}
  \rho_0^\ast(N{=}5) = 4.32~\mathrm{fm}\,,\qquad
  \rho_0^\ast(N{=}6) = 3.11~\mathrm{fm}\,.
  \label{eq:crossings}
\end{equation}
They are not the same, and no single common range describes both nuclei.
Figure~\ref{fig:levels} shows the two ends of the scan as level
schemes not to scale, each with the nucleus it reproduces and the one it misses.

Still, it is worth noting that 
the nucleus that is \emph{not} reproduced is described to within
$11\%$ ($^{20}$Ne at $\rho_0=3.0$~fm) or $16\%$ ($^{24}$Mg at
$\rho_0=4.3$~fm).

\section{A Hoyle-like resonance}
\label{sec:hoyle}

Nothing in the construction asks for a $3\alpha$ excited state. Nevertheless, once the
three-body force is switched on, $^{12}$C acquires a compact $0^+$ state just
above the $3\alpha$ threshold, in the position of the Hoyle state. It appears
only with the three-body force -- with $V_2$ alone the $N=3$ excited state is
unbound and pushed away by the Coulomb barrier (Table~\ref{tab:twobodyonly})
-- so it is driven by $V_3$, and it is not fitted.

\begin{table}[t]
  \caption{The $^{12}$C spectrum at the two ends of the scan, in MeV relative
  to the $3\alpha$ threshold. Only the $0^+_1$ ground state is fitted; the
  $2^+_1$ and the $0^+_2$ Hoyle analogue are predictions. Experimental values
  are $E_3$ of Table~\ref{tab:ikeda} for the ground state, the $2^+_1$ at
  $E_x=4.44$~MeV and the Hoyle state at $E_x=7.65$~MeV.}
  \label{tab:hoyle}
  \begin{ruledtabular}
  \begin{tabular}{lccc}
    & $\rho_0=4.3$~fm & $\rho_0=3.0$~fm & exp. \\
    \hline
    $0^+_1$ (fitted)   & $-7.270$ & $-7.269$ & $-7.27$  \\
    $2^+_1$            & $-1.462$ & $-0.721$ & $-2.83$  \\
    $0^+_2$ (Hoyle)    & $+0.160$ & $+0.359$ & $+0.379$ \\
  \end{tabular}
  \end{ruledtabular}
\end{table}

Table~\ref{tab:hoyle} collects the results. At $\rho_0=3.0$~fm -- the range
that reproduces $^{24}$Mg -- the resonance lands $0.359$~MeV above threshold
against $0.379$~MeV measured, a $5\%$ miss on a quantity that entered no fit.
At $\rho_0=4.3$~fm it is present but too low, at $0.160$~MeV. That the same
short range should get both $^{24}$Mg and the Hoyle state right is
suggestive, and it is the one place where the two ends of the scan are not
symmetric in quality.

A full characterization of this resonance -- its width, its monopole matrix
element $M(E0)$~\cite{yamada:2008_J.Phys.Conf.Ser.}, and the geometry of its wave
function~\cite{otsuka:2022_NatCommun} -- is beyond the scope of this paper and is
deferred to a separate study. What matters here is that it is present, that
it is in the right place, and that it was not asked for.

The $2^+_1$, by contrast, is badly reproduced at both ranges: $-1.46$ and
$-0.72$~MeV against $-2.83$~MeV. Its nature probably asks for
a description based either on the shape parameters or on an
angular-momentum-dependent potential~\cite{ali:1966_NuclearPhysics}.

\section{Conclusions}
\label{sec:conclusions}

When plotted on the universal straight-line version of the $N$-boson function,
the empirical Ikeda $\alpha$-cluster thresholds~\cite{ikeda:1968_Prog.Theor.Phys.Suppl.}
fall on it with remarkable precision. This linearity is the same one predicted
for bosons with a shallow 
state~\cite{gattobigio:2014_Phys.Rev.A,kievsky:2015_Phys.Rev.A,recchia:2022_Few-BodySyst}
and found in both zero-range EFT~\cite{bazak:2016_Phys.Rev.A} and Monte
Carlo~\cite{nicholson:2012_Phys.Rev.Lett.} calculations. No single member of a
one-parameter family of two-body potentials reproduces the line:
the $\alpha$-cluster ground states are not a two-body problem.

A minimal construction was proposed to fill the gap. The $\alpha$-$\alpha$
interaction is fixed inside the zero-shape theoretical framework~\cite{recchia:2022_Phys.Rev.A}
using a Bargmann potential~\cite{newton:1982_}
anchored to the $^{8}$Be resonance. The most economical three- and four-body
forces -- single Gaussians in the hyperradius, sharing a common range -- are
then fitted to two data points, $^{12}$C and $^{16}$O. Microscopic calculations for the
$N=5$ and $N=6$ systems yield parameter-free predictions: no single common range reproduces both
$^{20}$Ne and $^{24}$Mg exactly, but each is described to within $11\%$--$16\%$
of its experimental energy.

A Hoyle-like $0^+_2$ state emerges in $^{12}$C without being targeted by the fit.
At the range that reproduces $^{24}$Mg, the predicted resonance lies $5\%$ off
the measured position -- a non-trivial byproduct of a model designed for something
else. We report its emergence here, leaving a detailed study of its properties for future work.

The picture that emerges is that the $\alpha$-cluster ground states are
controlled by a universal law at the level of a single number per nucleus.  The
rest of the spectrum, which is quite complex, is not captured by
this simple picture. However, the deviations from exact universality encode genuine
many-body physics that a minimal force can capture to a useful degree.

\appendix

\begin{acknowledgments}
The author thanks A.~Kievsky for fruitful discussions.
He also gratefully acknowledges the Biblioteca degli Arconi in Perugia for
its generous hospitality and the inspiring working environment it provided
during the course of this research.
\end{acknowledgments}

\end{document}